# Symmetry Breaking of Counter-Propagating Light in a Nonlinear Resonator

**Authors:** L. Del Bino[1†], J. M. Silver[1†], S. L. Stebbings[1], P. Del'Haye[1]*

**Affiliations:**

[1]National Physical Laboratory (NPL), Teddington, TW11 0LW, United Kingdom.

*Corresponding author. Email: pascal.delhaye@npl.co.uk

†These authors contributed equally to this work

**Abstract**:

Light is generally expected to travel through isotropic media independent of its direction. This makes it challenging to develop non-reciprocal optical elements like optical diodes or circulators, which currently rely on magneto-optical effects and birefringent materials. Here we present measurements of non-reciprocal transmission and spontaneous symmetry breaking between counter-propagating light in dielectric microresonators. The symmetry breaking corresponds to a resonance frequency splitting that allows only one of two counter-propagating (but otherwise identical) light waves to circulate in the resonator. Equivalently, the symmetry breaking can be seen as the collapse of standing waves and transition to travelling waves within the resonator. We present theoretical calculations to show that the symmetry breaking is induced by Kerr-nonlinearity-mediated interaction between the counter-propagating light. This effect is expected to take place in any dielectric ring-resonator and might constitute one of the most fundamental ways to induce optical non-reciprocity. Our findings pave the way for a variety of applications including all optical switching, nonlinear-enhanced rotation sensing, optically controllable circulators and isolators, optical flip-flops for photonic memories as well as exceptionally sensitive power and refractive index sensors.

**Main Text:**

Bi-directional propagation of light is an important prerequisite for many types of optical elements including interferometers and resonators. At the same time, nonlinear optical effects have driven many photonic developments in the past decades. Interestingly however, *nonlinear optical interactions between counter-propagating light* remained largely unexplored. A possible reason is the difficulty in achieving the required optical power levels. In addition, many nonlinear optical processes like second/third harmonic generation and four-wave mixing are forbidden for counter-propagating pump light because of violation of momentum conservation. Here we present the observation of nonlinear interaction between counter-propagating light in an optical resonator. The interaction is mediated by the Kerr nonlinearity in an ultra-high-Q whispering gallery resonator, which provides the power enhancement to achieve the required light intensities. The nonlinear coupling between two counter-propagating light waves leads to a spontaneous symmetry breaking between the clockwise and counterclockwise circulating light above a certain threshold power. Such an effect has been theoretically predicted in the 1980s by Kaplan and Meystre in the context of nonlinear enhanced Sagnac interferometry for rotation sensing (*1*). Theoretical calculations show that the nonlinear interaction could enable significantly enhanced rotation sensors (*2, 3*) that are limited only by the relative power stability between the counter-propagating light. In addition the symmetry breaking could be used for precise optical power and refractive index sensing (*4*) and can play a critical role in the development of parity-time symmetric optical systems (*5*). Equally important, the interaction between counter-propagating waves leads to controllable non-reciprocal propagation of light. In particular, integrated photonic circuits require new ways for non-reciprocal light propagation that do not rely on magneto-optical effects (*6–13*). The presented results enable the development of novel types of integrated optical elements including optically controllable diodes, circulators and all-optical flip-flops (*14–17*). In our measurements we show a detailed analysis of the symmetry breaking, which manifests itself as a resonance splitting between clockwise and counterclockwise microresonator modes. In addition, we perform a threshold power analysis of the nonlinear interaction between counter-propagating light and demonstrate all-optical switching between clockwise and counterclockwise circulating light. Our experimental findings are in excellent agreement with theoretical calculations of Kerr-nonlinearity-induced symmetry breaking and non-reciprocal propagation of light.

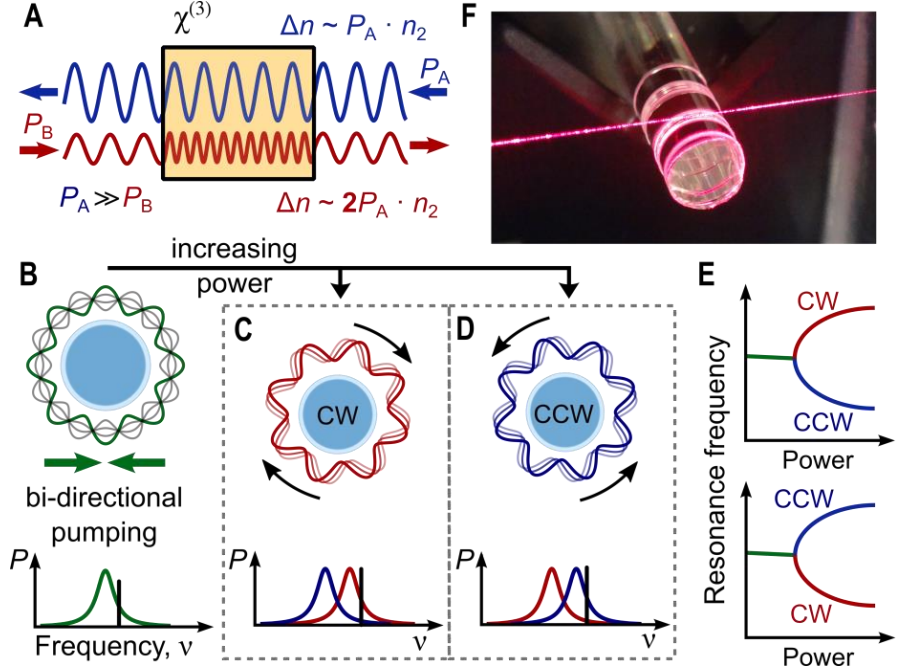

**Figure 1. Nonlinear interaction between counter-propagating light**. (**A**) Principle of Kerr-nonlinearity-mediated interaction between counter-propagating light without a resonator. Two counter-propagating but spatially overlapping light waves (offset in the diagram for clarity) with identical frequency will experience a different effective refractive index change $\Delta n$ depending on their powers ($P_A$, $P_B$). The light wave with lower optical power ($P_B$) experiences a stronger refractive index increase, which leads to a shorter wavelength. (**B**) Bi-directional pumping of a whispering gallery resonator at low power, generating a standing wave. When increasing the power, the system collapses either into state (**C**) with clockwise propagating (CW) light or state (**D**) with counterclockwise (CCW) propagating light. This symmetry breaking goes along with a resonance frequency splitting between the counter-propagating optical modes (shown in the lower part of panels B-D, where the black line denotes the pump frequency). (**E**) Symmetry breaking shown by the splitting between the CW and CCW resonance frequencies with increasing power. (**F**) Fused silica resonator and tapered fiber used in the experiments (highlighted with red laser light).

The nonlinear interaction between counter-propagating light in a nonlinear Kerr medium is illustrated in Figure 1A. Crucially, two counter-propagating light waves with equal wavelength *outside* of a nonlinear Kerr medium will have different wavelengths *within* the medium if their powers are unequal. This can be explained by cross-phase modulation between the two light waves, in which the weaker light wave experiences a stronger refractive index change. More specifically, the refractive index change $\Delta n$ induced by the nonlinear interaction is given by

$$\Delta n_A = \frac{n_2}{A_{\text{eff}}}(P_A + 2P_B) \quad \text{and} \quad \Delta n_B = \frac{n_2}{A_{\text{eff}}}(P_B + 2P_A), \tag{1}$$

with subscripts A, B indicating the two counter-propagating waves with powers $P_{A,B}$, $n_2$ being the nonlinear refractive index of the medium and $A_{eff}$ being the effective mode cross-section. It is important to note that a counter-propagating light wave induces twice the refractive index change compared to the change induced by self-phase-modulation (*18*, *19*). In the case of an optical resonator with $\chi^{(3)}$ nonlinearity, the difference in refractive index change leads to two different optical path lengths experienced by the counter-propagating modes. This is reflected in a splitting of the resonance frequencies of the clockwise (CW) and counterclockwise (CCW) modes as shown in Fig. 1C, D. Theoretical predictions suggest that such resonators exhibit symmetry breaking when simultaneously pumped with equal power in both the CW and CCW directions (*1*, *2*). This may be explained by a self-amplification of small power fluctuations between the counter-propagating light. With the laser frequency at higher frequency compared to the resonance, the optical mode with infinitesimally lower power experiences a stronger Kerr-shift and is pushed further away from the pump laser frequency. Simultaneously, the stronger mode experiences less Kerr-shift and moves towards the pump laser, such that it gains even more power. This increases the resonance splitting until the system comes to a new equilibrium and self-phase modulation does not allow the stronger mode to further approach the laser (similar to thermal self-locking (*20*) of microresonator modes to a laser). For low optical powers (Figure 1B), no splitting is induced and the CW and CCW circulating powers remain equal. Above a certain threshold power however, the state with equal coupled powers becomes unstable, and the system instead chooses one of the two states shown in Figures 1C and D, breaking the CW-CCW symmetry.

In quantitative terms, the optical powers that are coupled into the clockwise and counterclockwise circulating modes of a whispering gallery resonator are given by the following coupled equations:

$$P_{CW} = \frac{\eta P_{in,CW}}{1+\left(\frac{\delta}{\gamma}+\frac{1}{P_0}(P_{CW}+2P_{CCW})\right)^2} \quad (2)$$

$$P_{CCW} = \frac{\eta P_{in,CCW}}{1+\left(\frac{\delta}{\gamma}+\frac{1}{P_0}(P_{CCW}+2P_{CW})\right)^2} \quad (3)$$

Here, $P_{in,CW}$ and $P_{in,CCW}$ are the incident pump powers, $\delta$ is the detuning of the laser frequency with respect to the resonance without Kerr shift, $\gamma = \gamma_0 + \kappa$ is the loaded half-linewidth of the resonance for intrinsic and coupling-related decay rates $\gamma_0$ and $\kappa$, and $\eta = 4\kappa\gamma_0/\gamma^2$ is the coupling efficiency. The quantity $P_0 = \pi n_0 A_{\text{eff}}/(QF_0 n_2)$ is the coupled power at which nonlinear effects occur, with $n_0$ and $n_2$ being the nonlinear refractive indices for a resonator with effective cross-sectional area $A_{\text{eff}}$, a loaded quality factor $Q$, and intrinsic finesse $F_0$. Importantly, equations (2, 3) show that the nonlinear interaction with the counter-propagating light wave is twice as strong as the self-phase-modulation-induced interaction. This can be seen from the factor of two in the term $P_{\text{CW}} + 2P_{\text{CCW}}$ in equation (2). Moreover, a quick analysis of equations (2, 3) shows that for $P_{\text{in,CW}} = P_{\text{in,CCW}}$, symmetry breaking occurs when $\eta P_{\text{in,CW}}/P_0 > 8/(3\sqrt{3}) \approx 1.54$ over a range of $\delta/\gamma$ that depends on the value of $\eta P_{\text{in,CW}}/P_0$, and that at the threshold pump power, this range is limited to the single point $-5/\sqrt{3} \approx -2.89$.

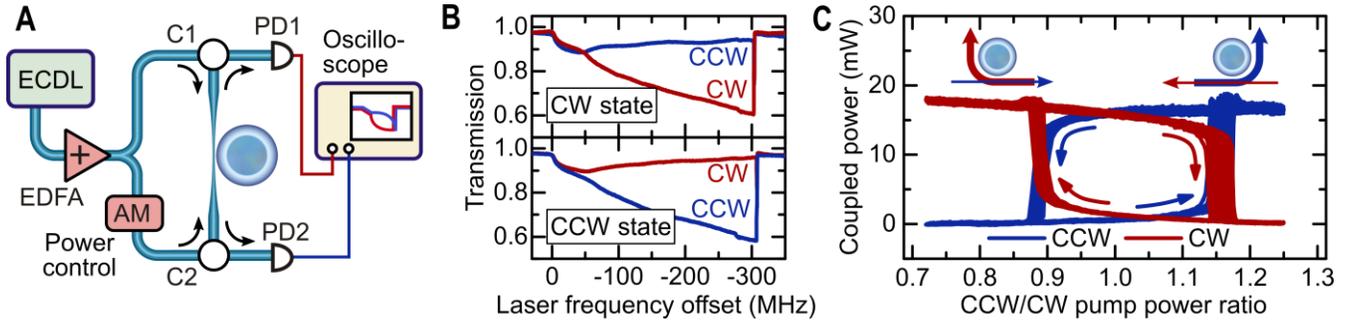

**Figure 2**. **Experimental setup and data showing nonlinearity-induced symmetry breaking between counter-propagating light.** (**A**) Schematic of the setup. The light of an external cavity diode laser (ECDL) is split into two parts and sent into a fused silica whispering gallery resonator from opposite directions. Two optical circulators C1, 2 split off the light coming from the microresonator and enable the measurement of the clockwise and counterclockwise transmission using photodiodes PD1 and PD2. The ratio between the counter-propagating pump powers can be adjusted using an amplitude modulator (AM). (**B**) Transmission vs. laser frequency measured on two consecutive laser sweeps across the resonance with equal powers. The system spontaneously "picks" one of two possible states. (**C**) Measurement of optical switching between the CW and CCW states, showing a hysteresis loop. The graph is obtained by tuning the laser into resonance and modulating the power in one direction up and down ~30 times. Inset: illustrations of the power flow in the two states.

Figure 2A shows our experimental setup, which is based on a 2.7 mm diameter high-Q whispering-gallery-mode fused silica microrod resonator (*21*). The resonator has a quality factor $Q_0$ of approximately $7 \times 10^7$ and an effective cross-sectional area $A_{\text{eff}}$ of approximately 60 $\mu$m$^2$ (see supplementary materials). Light from an amplified continuous wave external cavity diode laser (ECDL) in the 1.55 $\mu$m wavelength range is coupled into the resonator in both directions via a tapered optical fiber (*22*). A fiber-coupled electro-optic intensity modulator allows us to modulate the ratio of the pump powers $P_{in,CW}$ and $P_{in,CCW}$.

The spontaneous symmetry breaking is demonstrated in Figure 2B. Sweeping the laser frequency across the resonance with $P_{in,CW} = P_{in,CCW}$, the coupled powers $P_{CW}$ and $P_{CCW}$ initially follow the same trace before abruptly diverging, randomly choosing one of two states. Note that the total resonance shift of ~300 MHz also includes a thermally induced component (*20*, *23*), which however does not influence the splitting. An interesting feature of the nonlinearity induced symmetry breaking is hysteresis, which is shown in Figure 2C. Holding the laser frequency constant within the bistable regime, we modulate $P_{in,CCW}$ at 5 kHz by $\pm 25\%$. The observed hysteresis can be explained by the resonance frequency splitting between the two stable states. To overcome the resonance splitting and switch into the other state, the initially weaker pump direction has to be significantly increased to a power well above that of the counter-propagating pump light. Notably, this hysteresis allows the system to be used as an all-optical flip-flop or binary memory unit (*14*, *15*). Moreover, the non-reciprocal light propagation in each of the stable states can be exploited to realize optically switchable circulators or isolators.

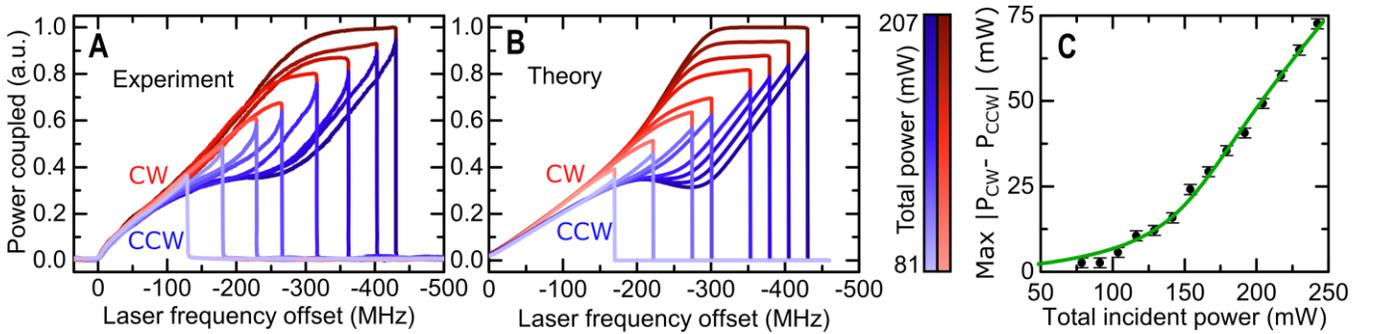

**Figure 3. Threshold and power dependence of the symmetry breaking between clockwise and counterclockwise propagating light.** (**A**) Amount of light coupled into the resonator as a function of the laser frequency for increasing incident powers. Higher powers (shown as darker colors) lead to an increased splitting between the clockwise (red) and counterclockwise (blue) modes. The incident power is 10% higher in the clockwise direction to avoid switching caused by fluctuations in the relative power. (**B**) Theoretical calculation of the power coupled into the resonator with the same parameters as in (A). (**C**) Maximum difference between clockwise and counterclockwise coupled powers as a function of the total input power. The black dots show the experimental measurements while the green curve is a theoretical fit for a pump power ratio of 0.9.

The amplification of power imbalances is tested against the theory in Figures 3A and B. Here the laser frequency is scanned across the resonance for a range of total pump powers, while keeping the ratio $P_{in,CCW}/P_{in,CW}$ fixed at 0.9. At low pump powers, no resonance frequency splitting is observed such that the two counter-propagating modes show the same profiles when sweeping across the resonance (lowest power curve in Figure 3A). Increasing the power, the effect of the cross-phase modulation induced Kerr shift leads to a large difference between the coupled powers. The theoretical results shown in Figure 3B use the same parameters as employed experimentally in Figure 3A. Included in the calculations is a thermally induced frequency shift (*20, 23*). A fit of the maximum coupled power difference vs. total pump power (Figure 3C) is in excellent agreement with the measured data. Small deviations between the curves in Figures 3A and B may be attributed to modelling the system using the assumption that the thermal shift of the resonance frequency is immediate, whereas in reality it has a delayed response (*23*).

The effects so far discussed are summarized by the theoretical plot in Figure 4A. The plot is calculated for $\eta P_{in,CW}/P_0 = 1.8$, slightly above the threshold for nonlinear mode splitting of 1.54. Outside the region of bistability demarcated by the gray lines, any pump power imbalance is amplified in the difference between the coupled powers. As the gray lines are approached, the gain factor diverges, rendering the system extremely sensitive to the precise pump powers and detunings as well as the losses in the resonator. This makes it well suited for applications in enhanced sensing of e.g. refractive index (*4*) or differential power. Notably, splitting in the resonance frequencies due to the Sagnac effect (*24*) is also amplified, enabling a novel form of optical gyroscope to be built (*1, 2*). These different regimes are demonstrated by the experimental data in Figure 4B, in which the pump power ratio is modulated about unity by ±3% during the frequency sweep (see supplementary materials). In the first region the coupled powers exhibit an enhancement of the oscillation that becomes stronger and stronger approaching the gray line. To the right of this line, the system responds to the oscillation by switching between the two stable solutions, crossing the unstable region instantaneously. In the region of maximum splitting between the CW and CCW mode, the power modulation is not sufficient to induce a switching between the states. This enables stable operation in either the CW or CCW state independent of optical power fluctuations. In the case of bidirectional pumping with equal powers we observe the splitting shown in the inset of Figure 4B, which is in good agreement with theoretical calculations (bold black curve in Figure 4A).

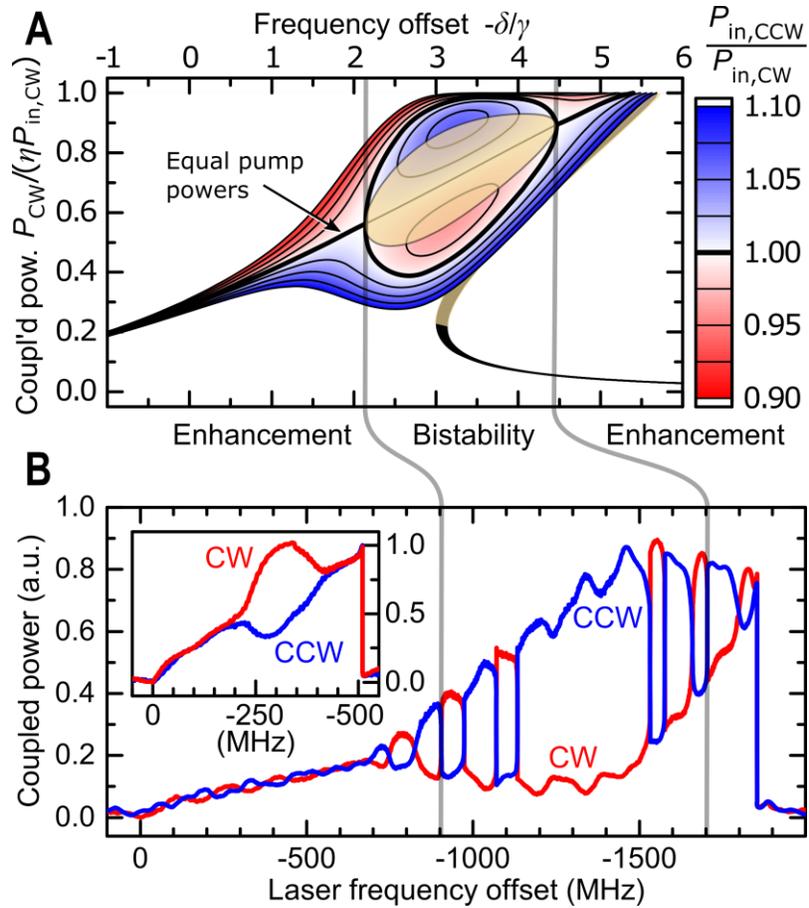

**Figure 4. Different regimes of nonlinearity-induced mode splitting.** (**A**) Theoretical prediction for the power $P_{CW}$ coupled into the resonator in the clockwise direction as a function of the normalized laser detuning frequency $\delta/\gamma$. The incident power imbalance is color-coded with red (blue) contours corresponding to higher power in the CW (CCW) direction. The bold black curve shows spontaneous symmetry breaking in the case of equal pump powers. In this region between the two gray lines, two stable solutions exist for a range of power imbalances. The beige shaded area contains unstable solutions. (**B**) Scan of the laser frequency across the resonance while backscattering-induced interference effects in the setup cause $P_{in,CCW}/P_{in,CW}$ to change about unity by ±3% at a period of ~170 MHz in the laser frequency. Outside the gray lines the coupled power difference is enhanced with respect to the incident power imbalance by the non-reciprocal Kerr shift. The enhancement increases approaching the bistable regime, in which the system begins to jump between the two stable configurations. The inset shows a sweep across a microresonator mode with equal power, leading to a "bubble"-shaped symmetry-broken region corresponding to the bold black trace in (A).

In conclusion, we demonstrate the observation of spontaneous symmetry breaking between counter-propagating light in a nonlinear resonator by pumping an optical microresonator equally in the clockwise and counterclockwise directions. Our results closely match our theoretical predictions based on nonlinear interaction between counter-propagating light induced by the Kerr nonlinearity. We show that above a threshold power, a symmetry-broken bistable regime exists as first predicted by Kaplan and Meystre (*1*, *2*). Several potentially far-reaching applications have been proposed for this effect. For example, the hysteresis manifested by the bistable regime allows the realization of an all-optical flip-flop (*14*, *15*). Furthermore, just below the threshold of bistability, the state of the system is exceptionally sensitive to minute imbalances between clockwise and clockwise circulating light (*1*). This enables a range of enhanced sensors for optical power, refractive index (*4*) and rotation (*1*, *3*), as well as new types of near-field probes. The observed nonlinearity-induced symmetry breaking and non-reciprocal propagation of light has significant potential for a variety of novel concepts in basic research and for advanced photonic applications.

**Acknowledgments:**

This work has been supported by the National Physical Laboratory Strategic Research Programme.

**References and Notes:**

1. A. E. Kaplan, P. Meystre, Enhancement of the Sagnac effect due to nonlinearly induced nonreciprocity. *Optics Letters*. **6**, 590–2 (1981).

2. A. E. Kaplan, P. Meystre, Directionally asymmetrical bistability in a symmetrically pumped nonlinear ring interferometer. *Optics Communications*. **40**, 229–32 (1982).

3. C. Wang, C. P. Search, Enhanced rotation sensing by nonlinear interactions in silicon microresonators. *Optics Letters*. **39**, 4376–4379 (2014).

4. C. Wang, C. P. Search, A Nonlinear Microresonator Refractive Index Sensor. *Journal of Lightwave Technology*. **33**, 4360–4366 (2015).

5. B. Peng *et al.*, Parity-time-symmetric whispering-gallery microcavities. *Nature Physics*. **10**, 394–8 (2014).

6. L. Bi *et al.*, On-chip optical isolation in monolithically integrated non-reciprocal optical resonators. *Nature Photonics*. **5**, 758–762 (2011).

7. L. Fan *et al.*, An all-silicon passive optical diode. *Science*. **335**, 447–450 (2012).

8. L. Feng *et al.*, Nonreciprocal Light Propagation in a Silicon Photonic Circuit. *Science*. **333**, 729–733 (2011).

9. F. D. M. Haldane, S. Raghu, Possible realization of directional optical waveguides in photonic crystals with broken time-reversal symmetry. *Physical Review Letters*. **100**, 013904 (2008).

10. N. Kono, K. Kakihara, K. Saitoh, M. Koshiba, Nonreciprocal microresonators for the miniaturization of optical waveguide isolators. *Optics express*. **15**, 7737–7751 (2007).

11. R. J. Potton, *Reports On Progress In Physics*, in press, doi:10.1088/0034-4885/67/5/R03.

12. F.-J. Shu, C.-L. Zou, X.-B. Zou, L. Yang, Chiral Symmetry Breaking in Micro-Ring Optical Cavity By Engineered Dissipation. *arXiv preprint arXiv:1604.08678* (2016).

13. Z. Yu, S. Fan, Complete optical isolation created by indirect interband photonic transitions. *Nature Photonics*. **3**, 91–94 (2009).

14. B. A. Daniel, G. P. Agrawal, Phase-Switched All-Optical Flip-Flops Using Two-Input Bistable Resonators. *Ieee Photonics Technology Letters*. **24**, 479–481 (2012).

15. M. Haelterman, All-optical set-reset flip-flop operation in the nonlinear Fabry-Perot interferometer. *Optics Communications*. **86**, 189–91 (1991).

16. L. Liu *et al.*, An ultra-small, low-power, all-optical flip-flop memory on a silicon chip. *Nature Photonics*. **4**, 182–187 (2010).

17. M. Soljacic *et al.*, Switching through symmetry breaking in coupled nonlinear micro-cavities. *Optics express*. **14**, 10678–10683 (2006).

18. R. Y. Chiao, P. L. Kelley, E. Garmire, Stimulated four-photon interaction and its influence on stimulated rayleigh-wing scattering. *Physical Review Letters*. **17**, 1158–1161 (1966).

19. A. E. Kaplan, Light-induced nonreciprocity, field invariants, and nonlinear eigenpolarizations. *Optics Letters*. **8**, 560–2 (1983).

20. T. Carmon, L. Yang, K. J. Vahala, Dynamical thermal behavior and thermal self-stability of microcavities. *Optics Express*. **12**, 4742–4750 (2004).

21. P. Del'Haye, S. A. Diddams, S. B. Papp, Laser-machined ultra-high-Q microrod resonators for nonlinear optics. *Applied Physics Letters*. **102**, 221119 (2013).

22. J. C. Knight, G. Cheung, F. Jacques, T. A. Birks, Phase-matched excitation of whispering-gallery-mode resonances by a fiber taper. *Optics Letters*. **22**, 1129–1131 (1997).

23. V. S. Ilchenko, M. L. Gorodetsky, Thermal Nonlinear Effects in Optical Whispering Gallery Microresonators. *Laser Physics*. **2**, 1004–1009 (1992).

24. E. J. Post, Sagnac effect. *Reviews of Modern Physics*. **39**, 475–493 (1967).